\documentclass[notitlepage,longbibliography,superscriptaddress,pre,nofootinbib]{revtex4-1}%
\usepackage{placeins}
\usepackage{pbox}
\usepackage{color}
\usepackage{relsize}
\usepackage{graphicx}
\usepackage[intlimits]{amsmath}
\usepackage{amsxtra, amssymb,fancyhdr, amsthm,latexsym}
\usepackage{cases}
\usepackage{verbatim}
\usepackage{float}
\usepackage[hidelinks]{hyperref}
\usepackage{textcomp}
\usepackage{epstopdf}
\usepackage[margin=0.75in]{geometry}
\usepackage{amsfonts}
\usepackage{amssymb}%
\providecommand{\U}[1]{\protect\rule{.1in}{.1in}}

\begin{document}
\begin{empty}

\title{Modelling of spatial infection spread through heterogeneous population: from
lattice to PDE models}
\author{Arvin Vaziry}
\author{T. Kolokolnikov}
\affiliation{Department of Mathematics and Statistics,
Dalhousie University Halifax,
Nova Scotia, B3H3J5, Canada}
\author{P. G. Kevrekidis}

\affiliation{Department of Mathematics and Statistics, University of Massachusetts,
Amherst, Massachusetts 01003-4515 USA}




\begin{abstract}
We present a simple model for the spread of an infection that incorporates
spatial variability in population density. Starting from first principle
considerations, we explore how a novel PDE with state-dependent diffusion can
be obtained. This model exhibits higher infection rates in the areas of higher
population density, a feature that we argue to be consistent with
epidemiological observations. The model also exhibits an infection wave whose
speed varies with population density. In addition, we demonstrate the possibility
that an infection can
\textquotedblleft jump\textquotedblright\ (i.e., tunnel) across areas of low
population density towards the areas of high population density. We briefly
touch upon the data reported for coronavirus spread in the Canadian province
of Nova Scotia as a case example with a number of qualitatively similar features as our
model. Lastly, we propose a number of generalizations of the model towards
future studies.

\end{abstract}
\maketitle
\end{empty}

\section{Introduction}

In the era of coronavirus, the ongoing public discussion frequently refers to
the reproduction number $R_{0}$, as a (simple) single-number diagnostic that
captures the entire epidemic for a given country or region; for a summary of
mathematical discussions of this diagnostic, we refer the interested reader
to~\cite{bailey,may,cc}. In reality, $R_{0}$ is a parameter which changes
locally, a feature that has not only been realized during the COVID-19
pandemic (see, e.g.,~\cite{arxiv1}), but indeed one that has been well-known
for similar outbreaks of other diseases such as dengue~\cite{ng}. For example,
it is natural to expect that areas with high population density and/or limited
public health measures are hit much harder than more rural areas, or regions
with strict health controls (masking and distancing). This suggests
the limited value of describing the entire population by a single reproduction number
$R_{0}.$ In light of such considerations, herein we are interested in
modelling of how the spread of disease depends on local
spatio-temporal circumstances. One of
the key parameter affecting the disease spread is population density. Our aim
is thus to develop a simple, potentially generalizable model which captures
the effect of population density and local differences on overall epidemic spread.

At the heart of many epidemiology models and in the frame of this study as
well, are the so-called compartmental models, consisting of various classes of
individuals and their interactions. Among the many possibilities that have
arisen not only in the context of COVID-19, but also earlier, we note the
formulation of ODE models \cite{hu2013scaling, bertozzi2020challenges,
kissler2020projecting, humphrey2020path}, statistical models
\cite{xia2004measles, bertozzi2020challenges}, stochastic models~\cite{stoch},
agent-based models \cite{mccallum2001should, ferguson2020report}, spatial
network models \cite{xia2004measles, colizza2008epidemic} and partial
differential equation (PDE) models \cite{Kevr, gai2020localized}; see also
\cite{chen2014modeling, mccallum2001should} for reviews. Some of these works
turn out to have a very deep influence on public thinking and policy
\cite{ferguson2020report, kissler2020projecting}.

The focus of the present work will be on spatially-distributed models
exploring the evolution of the infection not only temporally but also
spatially. Indeed, such models have a time-honored history, e.g., in the
format of meta-population models~\cite{COLIZZA2008450} and have been
extensively used in the context of COVID-19~\cite{review_meta}. Such models
have been used for a diverse host of countries including
China~\cite{timestr,substantial} and Spain~\cite{math8101677,arenas}, while a
comparison of different models developed, e.g., for the US can be found in the
so-called COVID-19 Forecast Hub\footnote{The relevant website is
\texttt{https://covid19forecasthub.org/doc/ensemble/}.}. On the other hand,
there exist also models that develop a PDE perspective such
as~\cite{viguerie2021simulating,Mammeri}, in addition to earlier work by the
present authors such as~\cite{Kevr,gai2020localized} (see also references
within these works).

Our aim in the present work is to complement the above approaches by means of
a first-principles look into the development of the interaction between the
different agents as they move through the spatial domain (and interact with
each other). In so doing, we will develop a nonlinear dynamical lattice based
approach, which can then be taken to the continuum limit, to yield a
systematic PDE model that can be more suitable towards the modeling of
COVID-19, as well as of other infectious diseases. Indeed, rather than
incorporating standard processes such as diffusion and advection into an ODE
SIR-type model, this perspective retrieves a nonlinear variant of diffusion
which seems to us to be more well-suited to such epidemic settings.
Additionally, a key advantage of the present model is that it enables a
variety of generalizations to account for effects of longer range interactions
(and, of course, additional effects such as those, e.g., of age distribution of
the pandemic impact). Such potential extensions will be highlighted along the
way. It is also relevant to mention that both for reasons of concreteness, but
also for practical ones related to the identifiability of the
model~\cite{plosone} (which does not escape us as a central issue and a consistent
source of concern about complex models), we opt within the present seed study to focus on
the prototypical SIR-type model. Generalizations to more detailed
models with a higher number of compartments will be evident, including
also in connection to earlier work of some of the authors~\cite{Kevr,rapti}.

Our presentation will be structured as follows. In section 2, we will present
the theoretical formulation of our model (and its potential extensions). In
section 3, we will use it to explore invasion waves and their respective
speed. In section 4, the onset of an infection outbreak will be examined.
Finally, after briefly touching upon the case example of Nova Scotia in
section 5, we conclude and present some future challenges in section 6.

\section{Theoretical formulation of the model}

We start with an agent based model, with the aim of deriving a cellular
automata model from it, and then consider its continuum limit to obtain a PDE
system. A similar procedure was used in \cite{short2008statistical} to derive
a spatio-temporal model of spreading of illegal activity. We assume that
individuals can get infected by going out of their home and traveling to new
locations. However they don't just simply walk at random, or diffuse:\ after
going out (e.g. say for shopping or work), they return to their original
(base) location.

To model individual motion, we discretize the space into bins. For
illustration (and although the procedure straightforwardly generalizes to
higher dimensions), we assume a one-dimensional grid whose bins are indexed by
$j=1\ldots N.$ Let $S_{j},I_{j},R_{j}$ denote the population of susceptible,
infected, and recovered in bin $j.$ As with the standard SIR\ model, we assume
that infection occurs with some probability $\beta$ per day when a susceptible
individual encounters an infected individual. A susceptible individual in bin
$j$ can get infected in two ways:\ they either get infected within its own bin
(e.g. infection spreading through families at home); or they might go out of
their home and get infected outside their bin (e.g., going to work, shopping,
etc); then return back to their original location. For simplicity, assume that
individuals travel only to neighbouring bins $j-1$ and $j+1$ for
work/shopping, then return back home. We will see how to extend the
model past this
simplifying assumption afterwards. In addition, assume for now that only
susceptible individuals can travel (we will deal with a more general case
below). Let $\alpha$ denote this daily travel rate (so that $\alpha S_{j}$
susceptibles travel from $j$ to $j+1$ and $\alpha S_{j}$ travel from $j$ to
$j-1$). Let $\Delta I_{j}$ denote new infections per day in bin $j$. With
above assumptions, we obtain,
\begin{equation}
\Delta I_{j}=\beta\left(  S_{j}-2\alpha S_{j}\right)  I_{j}+\beta\alpha
S_{j}I_{j-1}+\beta\alpha S_{j}I_{j+1}^{{}} \label{1713}%
\end{equation}

Here, $\beta\left(  S_{j}-2\alpha S_{j}\right)  I_{j}$ represents the daily
new infections that happen in bin $j;$ whereas $\beta\alpha S_{j}I_{j\pm1}$ is
the total number of new infections within bin $j$ acquired by individuals
going to work/shopping etc in the neighbouring bins, then returning home with
an infection (due to the interaction of these susceptibles with the infected
individuals in bins $j \pm1$.

The corresponding SIR\ model on a lattice then reads,
\[
S_{j}(t+1)=S_{j}-\Delta I_{j};\ \ \ \ I_{j}(t+1)=I_{j}+\Delta I_{j}-\gamma
I_{j},\ \ \ R_{j}(t+1)=R_{j}+\gamma I_{j}.
\]
We now consider the continuum limit of this model, in the limit of many bins.
Let $dx$ be the grid spacing, so that $I_{j}\approx I(x)$ where $x=jdx$. We
then estimate%
\[
\beta\left(  S_{j}-2\alpha S_{j}\right)  I_{j}+\beta\alpha S_{j}I_{j-1}%
+\beta\alpha S_{j}I_{j+1}\approx\beta SI+\beta\left(  dx\right)  ^{2}\alpha
SI_{xx}.
\]
and we estimate $S_{j}(t+1)-S_{j}(t)\approx S_{t}$ (up to a rescaling by the
time discretization increment $dt$ and similarly for $I$ and $R.$ The
resulting equations become%
\begin{equation}
S_{t}=-D\beta SI_{xx}-\beta SI,\ \ \ I_{t}=D\beta SI_{xx}+\beta SI-\gamma
I,\ \ \ R_{t}=\gamma I \label{cont}%
\end{equation}

where
\begin{equation}
D=\left(  dx\right)  ^{2}\alpha.
\end{equation}
Note that unlike many other PDE models \cite{holmes1994partial,
viguerie2021simulating, kevrekidis2020spatial, Mammeri}, the ``diffusion''
term depends explicitly on the susceptible population density $S(x,t).$
Moreover, the ``diffusion'' enters into equation for $S$ with a
\emph{negative} sign, whereas it has a positive sign in the equation for $I$.

Next, consider a more realistic model, where both susceptible as well
as (e.g., asymptomatic~\cite{review_meta,rapti})
infected individuals travel, with rates $\alpha_{S}$ and $\alpha_{I}$,
respectively.

Then (\ref{1713})\ gets replaced with%
\begin{align}
\Delta I_{j}  &  =\beta\left(  S_{j}-2\alpha_{S}S_{j}\right)  \left(
I_{j}+\alpha_{I}\left(  I_{j-1}+I_{j+1}-2I_{j}\right)  \right) \nonumber\\
&  +\beta\alpha_{S}S_{j}\left(  I_{j-1}+\alpha_{I}\left(  I_{j-2}%
+I_{j}-2I_{j-1}\right)  \right) \\
&  +\beta\alpha_{S}S_{j}\left(  I_{j+1}+\alpha_{I}\left(  I_{j+2}%
+I_{j}-2I_{j+1}\right)  \right)  .\nonumber
\end{align}
The limiting procedure results in equations (\ref{cont}), but with
$\alpha=\alpha_{S}+\alpha_{I}.$ Hence, we expect this to be the
prototypical PDE-type model within this class of compartmental systems.

The remainder of the paper is concerned with the study of continuum equations
(\ref{cont}). Before we do so, it is relevant to add a word about the
possibility that traveling does not solely occur to bin $j \pm1$ with rate
$\alpha\equiv\alpha_{1}$, but similarly to $j \pm2$ with rate $\alpha_{2}$
etc. Then, it is straightforward to show that the Laplacian term is replaced
by a nonlocal term of the form $S(x) \int K(x-y) I(y) dy$, where the (decaying
with distance) kernel $K$ is proportional to the probability of traveling
between locations of distance $|x-y|$. A straightforward Taylor expansion
around the vanishing argument of the kernel can be used to see that the
diffusivity $D$ above is proportional to the second moment (i.e., the
variance) of the above kernel. More specifically, assuming for simplicity an even
(or more generally isotropic) kernel
\begin{align}
\int K(x-y) I(y) dy = \int K(\xi) I(\xi+ x) d\xi\approx(\int K(\xi) d\xi) I(x)
+ D I_{xx} + \dots\label{extra1}%
\end{align}
Accordingly, the first term renormalizes $\beta$, while the second one
produces the diffusive approximation with $D=(1/2) \int K(\xi) \xi^{2} d\xi$.
We can thus see how such beyond nearest-neighbor terms can generalize the
model, while falling back to it in the simplest diffusive correction
level of approximation.
It is also interesting to further perceive how anisotropic kernels
may lead to directed (convective rather than diffusive) motion,
although the latter possibility will not be pursued further here.

\section{Examination of an invasion wave}

One of the main effects of introducing a spatial dimension, is that the
infection typically propagates from its origin. When the movement is
sufficiently slow, this propagation happens in a wave-like
fashion. One of the, arguably,
simplest settings exhibiting wave propagation is the context of KPP-Fisher
equation, modelling propagation of invasive species inside a favorable medium
(see, e.g.,~\cite{xin} for a review):%
\begin{equation}
u_{t}=du_{xx}+ru-su^{2}. \label{1516}%
\end{equation}

The travelling-wave solution has the form $u(x,t)=U(x-ct)$ where $U$ satisfies
the corresponding co-traveling ordinary differential equation (ODE)%
\[
-cU^{\prime}=dU^{\prime\prime}+rU-sU^{2}.
\]
We seek a wave propagating from left to right, so that $U(z)\rightarrow0$ as
$z\rightarrow+\infty,$ and $U\rightarrow r/s$ as $z\rightarrow-\infty.$
Following the relevant standard theory and linearizing at the front of the
wave ($z\rightarrow+\infty)$, we can seek a solution of the form
\[
U(z)\sim\exp\left(  -\lambda z\right)  ,\text{ as }z\rightarrow+\infty
\]
which yields a dispersion relationship between the speed $c$ and the decay
rate $\lambda$ of the form%
\begin{equation}
c=d\lambda+\frac{r}{\lambda}. \label{1515}%
\end{equation}
The minimum speed of propagation is obtained by minimizing (\ref{1515}) over
all admissable decay rates $\lambda>0,$ which yields%
\begin{equation}
c_{\min}=2\sqrt{dr}. \label{cmin}%
\end{equation}

\begin{figure}[ptb]
\includegraphics[width=0.98\textwidth]{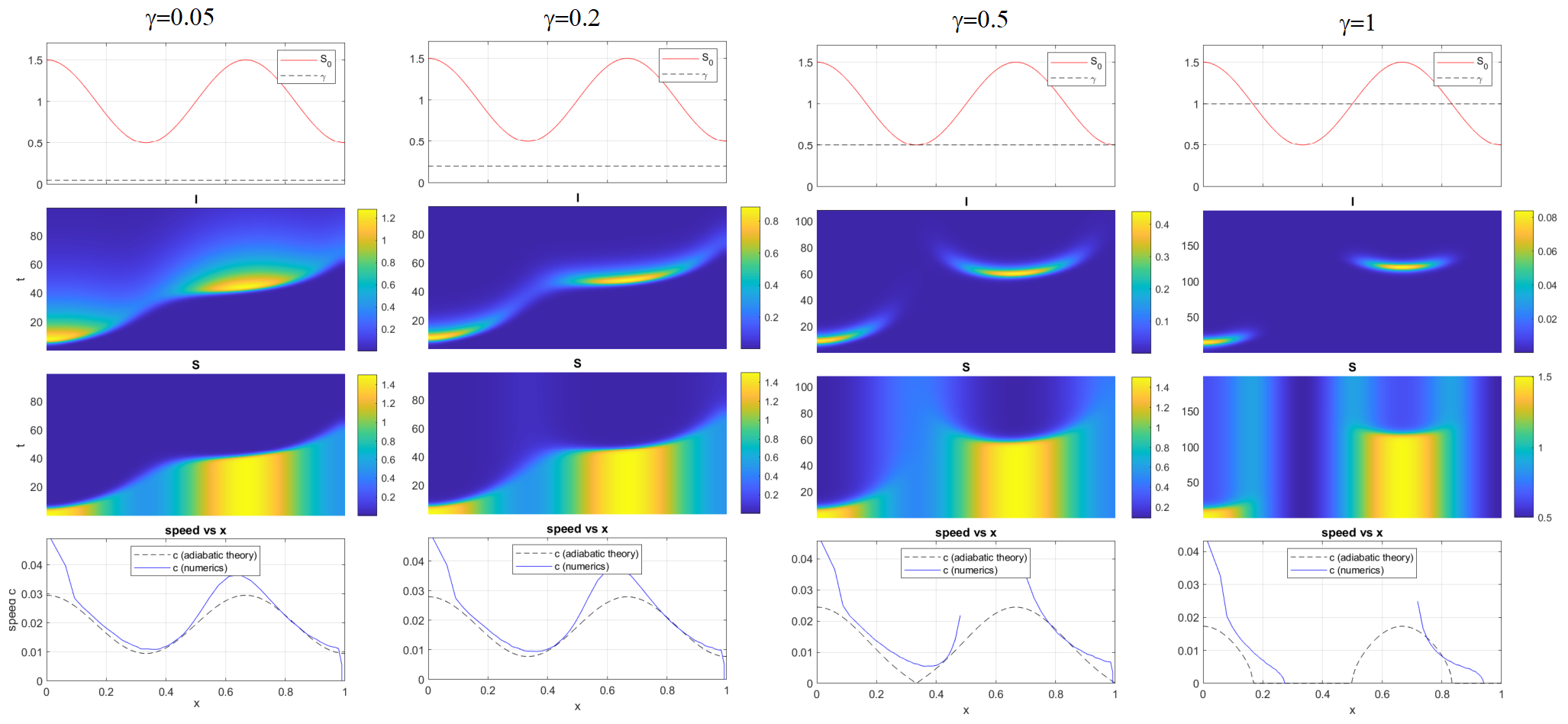}\caption{Simulation of an
infection wave propagating through a heterogeneous population, for several
value of $\gamma$ as indicated.\ Other parameters are: $\beta=1,\ \ S_{0}%
(x)=1+0.5\cos(3\pi x),$ $I_{0}(x)=0.01\exp\left(  -1000x\right)  ,$
and$\ D=0.0001.$ The top row shows $S_{0}$ and $\gamma$. Areas where $\beta
S_{0}(x)>\gamma$ (i.e. where red solid line is above the dashed line) are favorable
for outbreak. The second row shows $I(x,t),$ the infection density propagating
through the population. The third row shows $S(x,t),$ \ the density of
susceptibles. The last row shows the speed $c$ of the wave as a function of
wave position $x$, comparing numerics to the adiabatic theory (see text). Note
how the infection \textquotedblleft tunnels\textquotedblright\ through areas
of low infectivity in the last column. We used $N=200$ meshpoints and $\Delta
t=0.001.$ See text for the description of the numerical method.}%
\label{fig:wave}
\end{figure}
Numerical experiments confirm that the speed of propagation
approaches $c_{\min}$ for a wide range of initial conditions, so long as
$u(x,0)$ decays \textquotedblleft sufficiently fast\textquotedblright\ as
$x\rightarrow\infty.$ This is a well-known feature of the KPP-Fisher
equations~\cite{xin,murray}. Note that this speed only depends on linear terms
in (\ref{1516}) (i.e, it is independent of the value of $s$). Now suppose that
the parameters $d,r$ are functions of space $x.$ If they vary sufficiently
slowly, we expect that the speed of propagation will still be well
approximated by\ (\ref{cmin}). This is the so-called adiabatic approximation.
We now return to the SIR\ model of Eq.~(\ref{cont}). At the front of the
infection of wave, we estimate $S(x)$ by $S_{0}(x)$, where $S_{0}(x)$ is the
corresponding initial condition. The implicit assumption here is that $I,R\ll
S$ and hence maintaining $S \approx S_{0}$ is a reasonable approximation. Then, this
leads to the effective linear PDE for $I(x,t):$%
\begin{equation}
I_{t}\sim D\beta S_{0}(x)I_{xx}+(\beta S_{0}(x)-\gamma)I.\label{Ilin}%
\end{equation}
Assuming that the motion is sufficiently slow ($D\ll O(1)$), we linearize at
the front of the wave similarly to our discussion above for the KPP-Fisher
equation, and obtain the following approximation for the speed of propagation,%
\begin{equation}
c(x)\sim2\sqrt{D\beta S_{0}(x)(\beta S_{0}(x)-\gamma)}.\label{c}%
\end{equation}

Figure \ref{fig:wave}\ shows a comparison between the formula (\ref{c})\ and
full numerical simulations for several choices of $\gamma$. We used an
implicit-explicit finite difference scheme to simulate the PDE\ of
Eq.~(\ref{cont}). As can be seen in the Figure \ref{fig:wave}, the adiabatic
approximation (\ref{c}) works relatively well in the areas where $\beta
S_{0}(x)-\gamma>0.$ The formula breaks down in the areas where $\beta
S_{0}(x)-\gamma\leq0.$ These areas can be thought of \textquotedblleft buffer
zones\textquotedblright\ where effective infection growth is negative;
otherwise stated, the local $R_0$ is below unity and infection is
suppressed therein.
Nonetheless, the infection wave is able to \textquotedblleft tunnel
through\textquotedblright\ these areas, with some delay; see Section 5 for
further investigation of this phenomenon.

\section{The onset of the outbreak}

Note that equations (\ref{cont}) admit a \textquotedblleft
trivial\textquotedblright\ solution corresponding to no outbreak; namely
$I(x,t)=0$ and $S(x,t)=S_{0}(x)$ where $S_{0}(x)$ describes the initial
population distribution. We now explore the conditions for the initiation of
the outbreak. At the onset of the outbreak, we may assume that $I(x,t)\ll1$.
Linearizing Eq.~(\ref{cont}), in analogy to what is done for the ODE variant
of the model to obtain the bifurcation associated with the spreading of the
infection~\cite{bailey,may,cc}, leads to equation for $I$ only of the form
(\ref{Ilin}). Looking for solutions of the form $I\left(  x,t\right)
=e^{\lambda t}\phi(x)$, we obtain an eigenvalue problem%
\begin{equation}
  \frac{\lambda+\gamma}{\beta S_{0}(x)}\phi=D\phi_{xx}+\phi.
  \label{eig}%
\end{equation}
First, consider the limit $D=0.$ In this case, each point $x$ in space evolves
separately, and the eigenvalues $\lambda$ are given by $\lambda\sim\beta
S_{0}(x)-\gamma.$ The outbreak is therefore prevented when $\beta
S_{0}(x)<\gamma$ for all $x,$ or $\gamma>\gamma_{c}\,,$ where
\begin{equation}
\gamma_{c}=\beta\max_{x}S_{0}(x). \label{1729}%
\end{equation}
This can be thought of as a ``spatially extended'' generalization of the ODE
result, in that the points in space are practically independent, hence for the
epidemic to be suppressed, this needs to be achieved ``individually'' for
every spatial point.

More generally, we define $\gamma_{c}$ to be a threshold value of the decay
parameter $\gamma$, corresponding to the zero-eigenvalue of (\ref{eig}).
Namely, $\gamma_{c}$ satisfies%
\begin{equation}
\frac{\gamma_{c}}{\beta S_{0}(x)}\phi=D\phi_{xx}+\phi; \label{1731}%
\end{equation}
the outbreak occurs if and only if $\gamma<\gamma_{c}$. For general $S_{0}(x)$
and $D,$ the problem (\ref{1731})\ does not have an explicit solution. However
we expect $\gamma_{c}$ to approach (\ref{1729})\ as $D\rightarrow0.$ We now
derive the corrections to (\ref{1729})\ in the limit of small but non-zero
$D$, i.e., for $0<D\ll1$ using asymptotic analysis. We expect the outbreak to
first occur near the maximum of $S_{0}$. Let $x_{m}$ be the point at which
$S_{0}(x)$ has its maximum. As such, we expand:%
\[
x=x_{m}+\varepsilon y,
\]
where $\varepsilon$ is a small constant to be determined. Near $x_{m},$ write:%
\[
S_{0}(x)\sim A\left(  1-B\varepsilon^{2}y^{2}\right)  +O(\varepsilon
^{3})\text{, \ \ where }A=S_{0}(x_{m});\text{ \ }AB=-S_{0}^{\prime\prime
}(x_{m})/2.
\]
and we expand $1/S_{0}(x)\sim\left(  1+B\varepsilon^{2}y^{2}\right)  /A.$
Problem (\ref{1731})\ then becomes%
\[
\frac{\gamma_{c}}{A\beta}\left(  1+B\varepsilon^{2}y^{2}\right)  \phi\sim
D\varepsilon^{-2}\phi_{yy}+\phi
\]
We now choose $\varepsilon$ so that $B\varepsilon^{2}=D\varepsilon^{-2}.$ In
other words we let
\[
\varepsilon:=D^{1/4}B^{-1/4}.
\]
Assuming $\varepsilon$ is small, to leading order we obtain an eigenvalue
problem%
\begin{equation}
\phi_{yy}-y^{2}\phi=\mu\phi,\text{ \ \ }y\in%
\mathbb{R}
\label{quantum}%
\end{equation}
with
\begin{equation}
\mu=\left(  \frac{\gamma_{c}}{A\beta}-1\right)  D^{-1/2}B^{-1/2}. \label{1018}%
\end{equation}
Equation (\ref{quantum}) is a well-known quantum-harmonic oscillator
eigenvalue problem whose eigenfunctions are given in terms of Hermite
polynomials multiplied by a Gaussian.\ The corresponding eigenvalues are given
by%
\[
\mu=1,3,5,7,\ldots
\]
The smallest eigenvalue is $\mu=1.$ Setting $\mu=1$ in (\ref{1018})\ we obtain
the following formula for the threshold value $\gamma_{c}:$%
\begin{equation}
\frac{\gamma_{c}}{\beta}\sim S_{0}(x_{m})-D^{1/2}\left(  -S^{\prime\prime
}(x_{m})/2\right)  ^{1/2}S_{0}(x_{m})^{1/2}+O(D). \label{gammac-smallD}%
\end{equation}
For example, take $S_{0}(x)=a+\sin\left(  \pi x\right)  ;$ $\beta
=1,\ \ x\in\left(  0,1\right)  .$ Then the maximum occurs at $x_{m}=0.5$ and
we obtain%
\begin{equation}
\gamma_{c}\sim1+a-D^{1/2}\pi\left(  1+a\right)  ^{1/2}2^{-1/2} \label{1733}%
\end{equation}
The following table compares the formula (\ref{1733}) with the fully numerical
solution of the eigenvalue problem (\ref{1731}), in the case of $a=0:$%

\[%
\begin{tabular}
[c]{lllll}%
$D$ & $0.01$ & $0.005$ & 0.0025 & 0.00125\\
$\gamma_{c}$ from numerics (\ref{1731}) & 0.7686 & 0.8429 & 0.8889 & 0.9214\\
$\gamma_{c}$ from asymptotics (\ref{1733}) & 0.7778 & 0.8389 & 0.8871 &
0.9206\\
Relative error & 1.18\% & 0.47\% & 0.20\% & 0.093\%
\end{tabular}
\ \
\]
The relative error appears to scale with a direct proportionality to $D$.

Let us also study the asymptotics in the limit of large $D$, on the domain
$x\in\left[  0,L\right]  $ with Neumann boundary conditions $\phi\left(
0\right)  =\phi(L)=0.$ In this case, we expand $\phi$ in (\ref{1731})\ as:%
\[
\phi=\phi_{0}+\frac{1}{D}\phi_{1}+\ldots
\]
At leading order in $D,$ we obtain $\phi_{0xx}=0.$ Together with boundary
conditions $\phi^{\prime}(0)=\phi^{\prime}(L)=0,$ this yields $\phi_{0}%
(x)=$const. By scaling, we may then take $\phi_{0}=1.$ The next-order equation
for $\phi_{1}$ then becomes%
\begin{equation}
\frac{\gamma_{c}}{\beta S_{0}(x)}=\phi_{1xx}+1. \label{1928}%
\end{equation}
We then integrate both sides from to obtain:%
\begin{equation}
\gamma_{c}\sim\beta\left(  \frac{1}{L}\int_{0}^{L}\frac{1}{S_{0}(x)}\right)
^{-1},\ \ \ D\gg O(1) \label{gammac-largeD}%
\end{equation}
The quantity\ $\left(  \frac{1}{L}\int_{0}^{L}\left(  S_{0}(x)\right)
^{-1}\right)  ^{-1}$ is called the harmonic average of $S_{0}(x).$

For example, take $S_{0}(x)=a+\sin\left(  \pi x\right)  $ with $x\in\left(
0,1\right)  .$ Then (\ref{1928}) integrates to%
\begin{equation}
\gamma_{c}\sim\left\{
\begin{array}
[c]{c}%
\frac{\pi\sqrt{1-a^{2}}}{\log\left(  1+\sqrt{1-a^{2}}\right)  -\log\left(
1-\sqrt{1-a^{2}}\right)  },\ \ \ 0<a<1\\
\pi/2,\ \ \ \ \ a=1\\
\frac{\pi\sqrt{a^{2}-1}}{\pi-2\arctan\left(  \left(  a^{2}-1\right)
^{-1/2}\right)  },\ \ a>1
\end{array}
\right.  \label{largeD}%
\end{equation}

\begin{figure}[tb]
\includegraphics[width=0.48\textwidth]{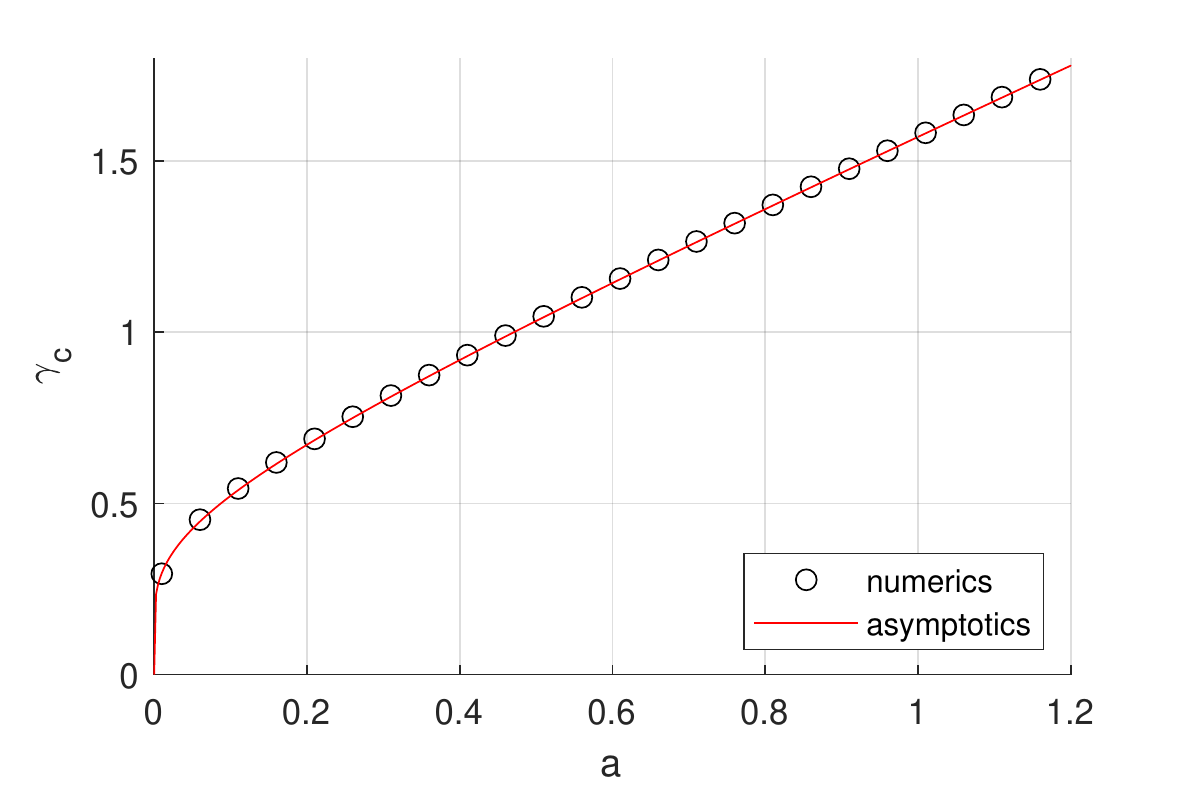}\includegraphics[width=0.48\textwidth]{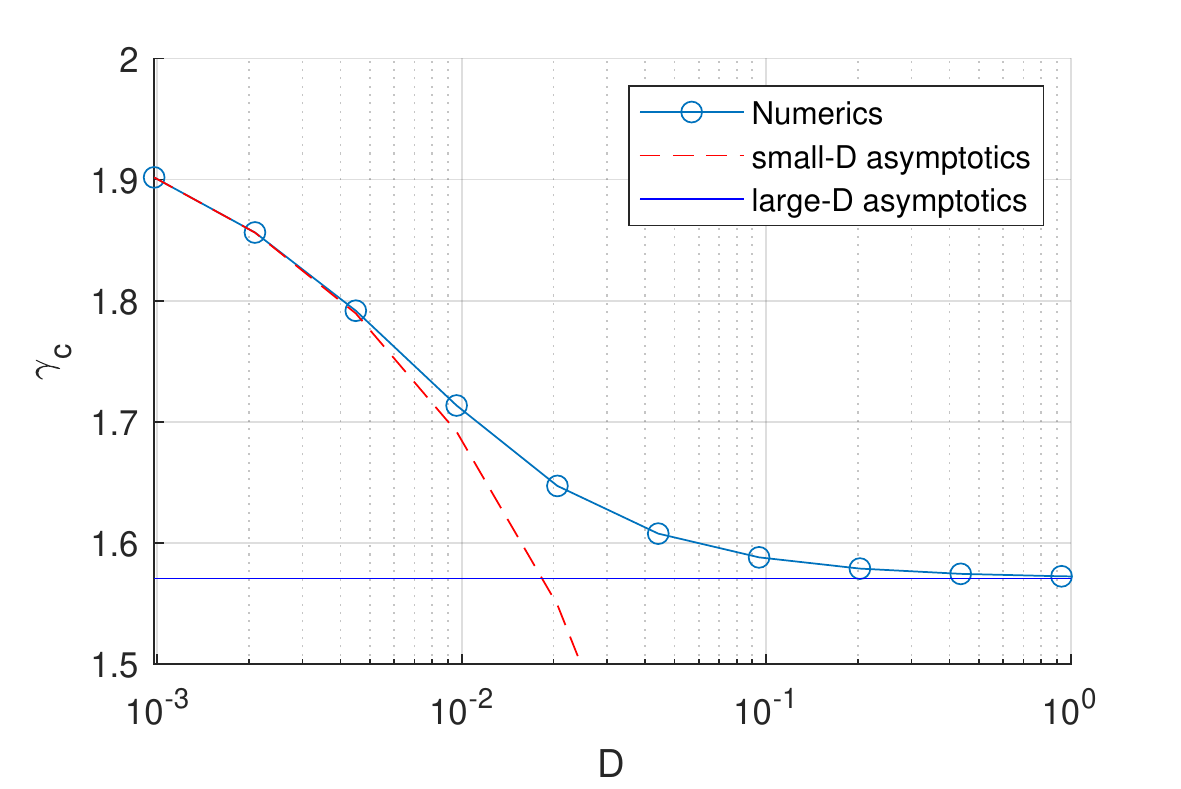}\caption{Left:\ Threshold
for outbreak $\gamma_{c}$ in the limit of \textquotedblleft
large\textquotedblright\ $D.$ Here, $D=1$ and $S_{0}(x)=a+\sin(\pi x)$,
$x\in\left(  0,1\right)  ;$ $\beta=1.$ The numerical solution of (\ref{eig})
and asymptotics given by (\ref{largeD}) are both shown. They are
indistinguishable, with relative error less than 0.1\%. Right:\ Threshold as a
function of $D$ with $S_{0}(x)=1+\sin(\pi x).$ Small and large-$D$
asymptotics are also shown.}%
\label{fig:gamma}%
\end{figure}

Figure \ref{fig:gamma}(left) compares the asymptotics (\ref{largeD}) with full
numerical simulations of (\ref{1731}) for a wide range of $a$, and with $D=1.$
Despite a relatively small value of $D,$ the agreement is excellent over the
entire range of $a$ (within 0.1\%). On the right, we fix $a=1$ and vary $D;$
as can be seen, both large- and small- $D$ asymptotics agree very well with
full numerics. The intermediate regime of $D$, where neither of our
approximations
is value illustrates the most substantial deviations, yet we still
have a very adequate description of the two asymptotic limits.

Finally, note that for constant population density $S_{0},$ the theshold
$\gamma_{c}$ defined by (\ref{1731})\ is independent of $D$, and both
(\ref{gammac-largeD})\ and (\ref{gammac-smallD}) yield $\gamma_{c}=\beta
S_{0}.$ This may also be rather natural to expect as in that case, the
diffusion term is ``deactivated'' and we are effectively back to the ODE
problem case. One might naively expect that in the large-$D$ limit, $S_{0}$
would be replaced by the arithmetic average of $S_{0}(x).$ However our
analysis shows that the more appropriate formula is to take a \emph{harmonic}
average of $S_{0}(x)$ as in (\ref{gammac-largeD}).

\section{Indicative observations from COVID-19 in Nova Scotia and
\textquotedblleft tunneling\textquotedblright}

\begin{figure}[tb]
\includegraphics[width=0.98\textwidth]{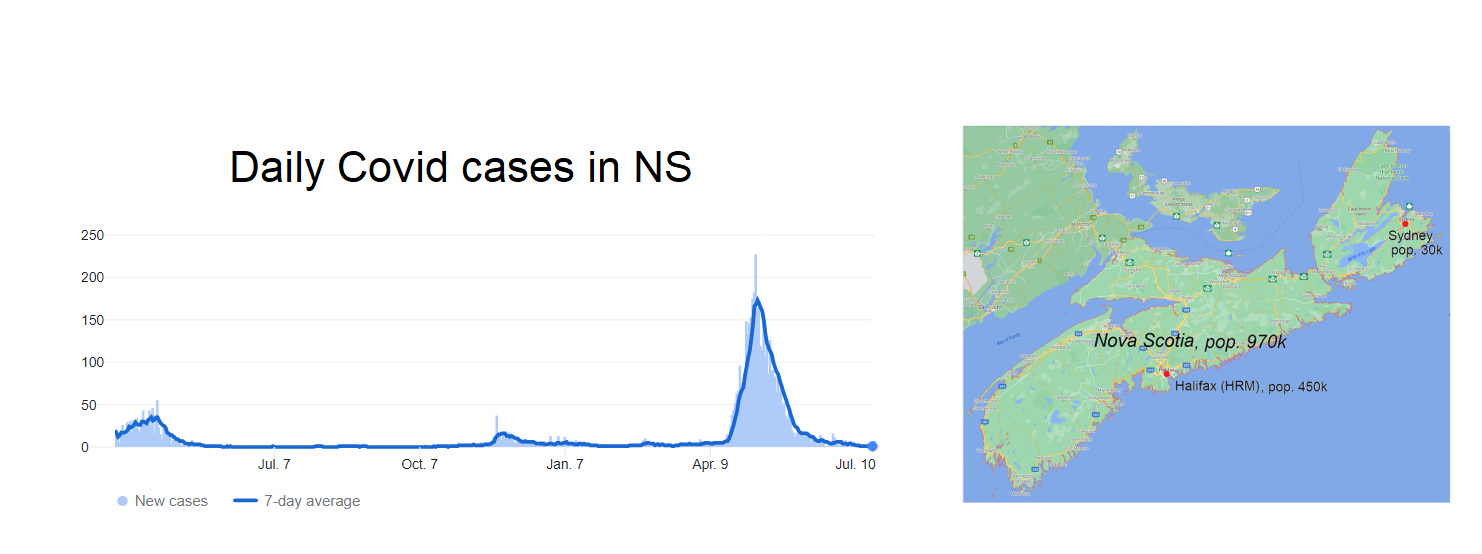}\caption{Daily COVID-19 cases for
the Province of Nova Scotia. Around 80\% of the cases occurred in the Halifax
Regional Minicipality, which contains about 50\%\ of the population of Nova
Scotia.}%
\label{fig:ns}%
\end{figure}

As a case study, consider the Canadian province of Nova Scotia where some of
the authors of this paper reside. It has a population of about 1 million, with
slightly less than half of those living in Halifax Regional Minicipality
(HRM:\ the city of Halifax and surrounding area). The second-biggest town is
Sydney (see map) with a population of 30,000. Much of the rest of the province
has relatively low population density. Nova Scotia managed to completely
suppress the initial outbreak in the spring of 2020 using very strict
stay-at-home orders and border controls. Any visitor required a strict
self-isolation quarantine of 2 weeks upon entry. As a result, there were very
few locally-transmitted cases up until April 2021; stringent health measures
managed to extinguish the few localized outbreaks that did occur before they spread.

Figure \ref{fig:ns} shows the daily COVID\ case numbers for Nova Scotia. In
total, as of July 2021, Nova Scotia had about 5800 cases, which is about 0.6\%
of the total population of 1 million. About 70\% of these cases occurred during
the \textquotedblleft third wave\textquotedblright\ in April-June, 2021. Very
few cases occurred in-between the three waves -- and most of those were
travel-related in quarantine (i.e., not involving community spread). Although
less than half of NS\ population lives in HRM, it was responsible for 79\% of
the cases overall, and 81\% of the cases in the third wave. Another 10.5\% of
cases occurred in Sydney, about 400km (4.5 hours drive)\ from Halifax, having a
population of 30,000. Together, HRM\ and Sydney were responsible for over 90\%
of all infections, despite having about half of the overall population of the
province. Despite its relatively smaller size, the infection rate in Sydney
was about 2.5 times that of Halifax during the third wave.

The main takeaway lesson from this brief data summary, in connection to
the qualitative model features discussed herein, is that the rate of infection is much higher in denser urban
regions than the rest of Nova Scotia, which is mainly rural with low
population density. This is indeed consistent with our model and its corresponding
observations. In addition, due to stringent health measures, it is likely that
the epidemic in most of the regions of Nova Scotia did not spread -- even
during the third peak -- as almost all infections came from HRM\ and Sydney --
the two biggest population centers in Nova Scotia.
Despite strict travel restrictions (even inter-provincial travel was banned
during the third wave in May 2021), the infection was able to ``tunnel
through"\ the rural areas from HRM\ to Sydney.\footnote{It is also interesting
to note that there are other significant population centers closer to HRM that
\emph{did not} see anything near the size of outbreak in Sydney. This includes
the towns of Truro (pop. 23000, one hour drive from Halifax)\ and New Glasgow
(pop. 19000, 2 hours drive from Halifax) that did not see any significant
outbreaks during the third wave. The outbreak in Sydney started with a hockey
game, when kids and families from Halifax visited Syndey for a hockey
tournament at the onset of the third wave, a potential superspreader event. At
the end of the day, our simple model is insufficient to make predictions at
such localized detail; much of the outbreaks are driven by random events and
the luck of the draw, which our deterministic model is not designed in
this first installment thereof to deal with. This is naturally an
intriguing challenge for further work.}

\begin{figure}[tb]
\includegraphics[width=0.49\textwidth]{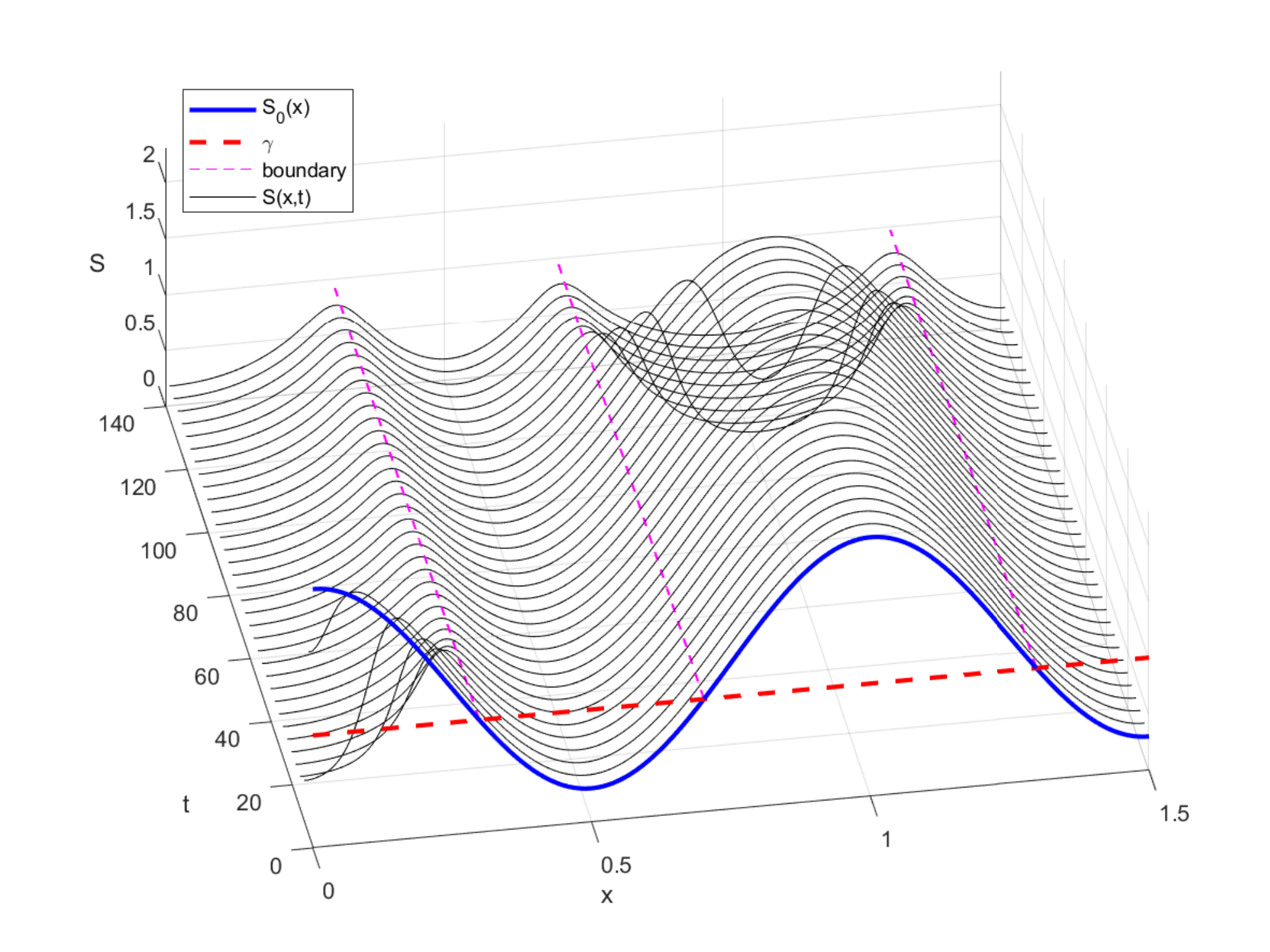}
\includegraphics[width=0.49\textwidth]{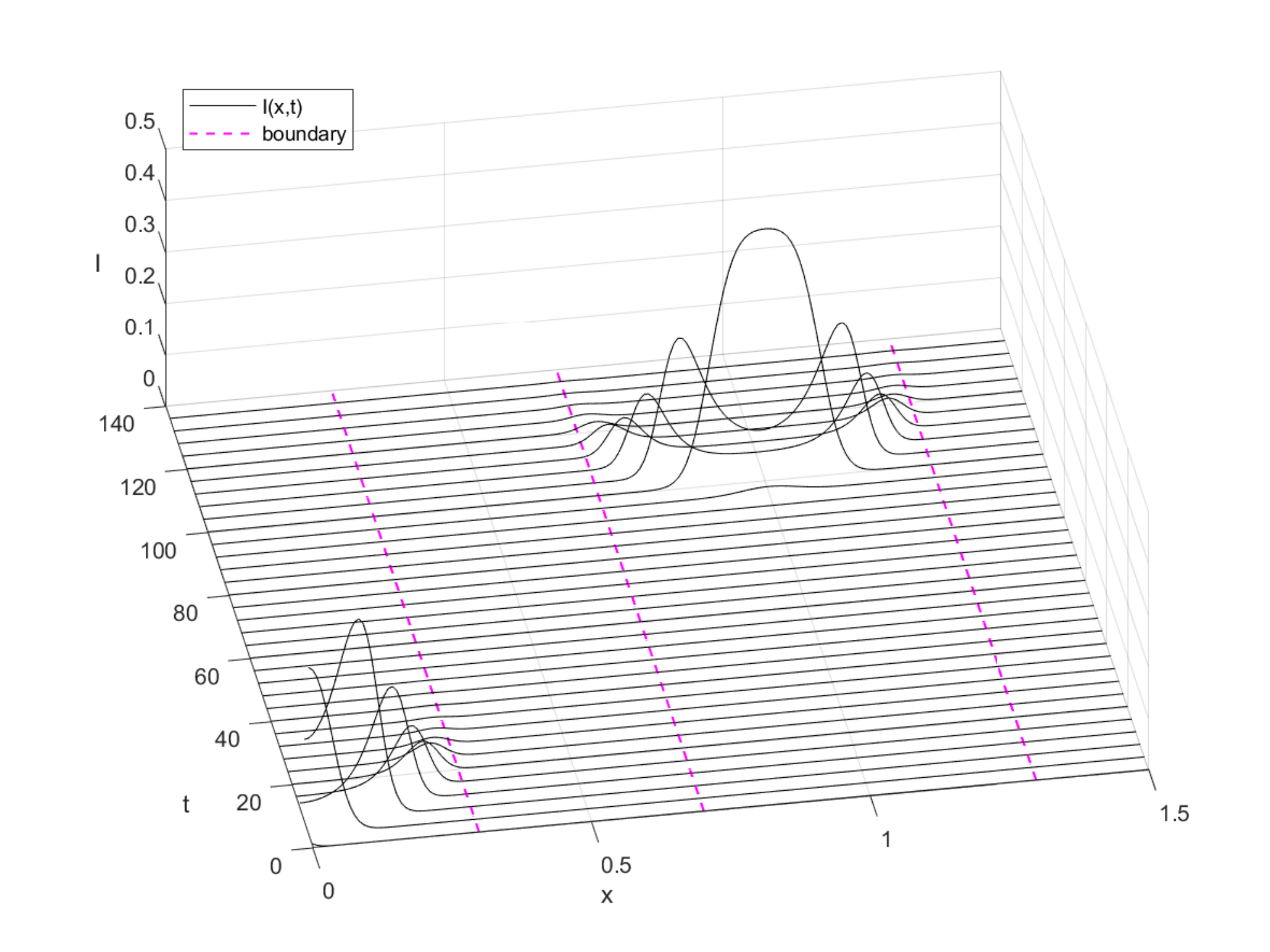}\caption{Infection
\textquotedblleft tunneling\textquotedblright\ through a barrier. Initial
conditions was taken to be $S_{0}(x)=1.3+\cos(2\pi x)$ with $\gamma=\beta=1$
and $x\in\lbrack0,1.5].$ Without spatial interactions ($D=0$), the disease is
suppressed in the middle region $x\in\left[  0.298,0.701\right]  $ as well for
$x>1.298.$ Here, we take $D=0.00005.$ The disease is introduced at $t=0$ at
the left end $x=0$; corresponding to initial conditions $I(x,0)=0.001\exp
(-1000\ast x)$. An infection wave propagating to the right is initially
observed, but appears to die out around $t\approx30$ as it hits the
buffer region  at
$x\approx0.3$. However it is able to \textquotedblleft tunnel
through\textquotedblright\ the buffer region, re-appearing at $x=1$ (where
$S_{0}$ has its maximum) when $t\approx90$, then propagating from there to the
rest of the infectious region $x\in\lbrack0.7,1.3].$}%
\label{fig:tunnel}%
\end{figure}

Motivated by the above observations, we now show that our model can reporoduce, at least qualitatively, a
\textquotedblleft tunneling-through\textquotedblright\ effect, where the
infection can spread between two regions of locally positive growth, even when
separated by a \textquotedblleft buffer zone\textquotedblright\ of negative
growth (i.e., infection suppression). Consider a sample simulation as shown in Figure \ref{fig:tunnel}, with
$S_{0}=S_{0}(x)=1.3+\cos(2\pi x)$ with$\ x\in(0,1.5)$ and $\beta=\gamma=1.$
Locally (in the limit of $D=0$), the infection is suppressed in the middle
region $x\in\left(  0.298,0.701\right)  $ as well as for $x>1.298$ where
$S_{0}(x)\beta<\gamma,$ and grows to the left and to the right of that region.
We initially introduce the infection near the left boundary of $x=0.$ The outbreak then
takes over the entre left region $0\leq x\leq0.298$ by the time $t=20.$ Then
for a relatively long time $20<t<100,$ nothing appears to happen. But
eventually at around $t\approx100,$ the infection manages to \textquotedblleft
jump\textquotedblright\ over to the right region and re-appears at $x=1$
(where $S_{0}(x)$ has its maximum), then spreads from there both to the left
and to the right until the entire region $0.701\leq x\leq1.298$ is infected.
It is interesting to note that when the infection re-appears at $t\approx100,$
it does so at $x=1$ rather than $x\approx0.7.$ The reason merits further
investigation, but roughly speaking, this happens because the local growth
rate of infection is given roughly by $S_{0}(x)\beta-\gamma$, and is the
highest at the maximum of $S_{0}(x).$

\section{Conclusions and Future Work}

We have presented a model of spatio-temporal infection spread. We have started
from a lattice variant of the problem and considered a first-principles
inclusion of mobility according to which people move to new, adjacent locations (for
work, shopping or other purposes), get infected and return to their base in
that new infected state. The model allows for extensions whereby the mobility
is to different locations (rather than to adjacent bins) with a presumably
decaying over distance kernel. The latter constitutes an interesting 
variant of the current model relevant to examine in future work. Considering the continuum limit of the
considered cellular automaton, we obtained a PDE (\ref{cont})\ with
state-dependent diffusion terms. Essentially, the scope of our work is to
advocate the relevance of consideration of such terms, in addition to local
ones and, arguably, instead of regular diffusion processes in this setting.
The key assumption in our modelling is that while individuals move around,
they don't diffuse, while infection does. While numerous PDE\ models
exist in epidemiology (see, e.g., \cite{Kevr, holmes1994partial,
gai2020localized, viguerie2021simulating, Mammeri} for a sample), most assume
either constant diffusion, or diffusion that is prescribed to be spatially-dependent.
By contrast, we present a first-principles derivation of Eq.~(\ref{cont}%
)\ from the underlying cellular automata representation of the basic infection
mechanisms. Our model naturally leads to a diffusion that scales with
the current
number of susceptibles.

Introducing a spatial component to a basic SIR\ model spread also naturally
explains why areas of high population density experience higher infection
rates than more rural areas (for related approaches see e.g.
\cite{hu2013scaling, kolokolnikov2021law}). We also generalized the concept of
the reproduction number in this spatially variable setting, by deriving an
eigenvalue problem (\ref{eig})\ whose solution describes overall decay or
spread of the disease. Importantly, the relevant eigenvalue problem
near the maximum of the susceptible population can be
approximated by a quantum harmonic oscillator which allows an approximate
analytical expression for the critical clearance rate that would avoid the
spreading of infection. We have tested the relevant predictions numerically,
finding very good agreement with our theoretical results, where appropriate.

Aside from spatially-dependent infection rates, our model demonstrates the
difficulty of suppressing the outbreaks. As illustrated in Figure
\ref{fig:tunnel},\ the disease can \textquotedblleft tunnel\textquotedblright%
\ between \textquotedblleft islands\textquotedblright\ of positive growth
separated by areas of negative growth (i.e., decay) of the epidemic. A
better
understanding and more systematic quantification of such phenomena is planned
for future work.

There are also numerous additional dimensions in which the present
consideration can be extended (both literally and figuratively). Indeed, here
we restricted considerations to one-dimensional settings, i.e., ``geographic
corridors''. In line with other works such as~\cite{Kevr,
viguerie2021simulating}, it is naturally more relevant to explore
two-dimensional domains. In addition, it is of substantial interest to
consider infections across different age groups. Our considerations herein
have assumed that the infectiousness and especially recovery properties of the
entire population are the same, however it is well-understood that COVID-19
has a far more severe impact on more senior individuals with weakened immune
system; indeed, this has been been the basis for designing relevant non-pharmaceutical
intervention strategies~\cite{fokas1}.
It is then of interest to introduce kernels of interaction across a
``synthetic dimension'' representing age (in addition to spatial dimensions).
There, interactions are predominant along the ``diagonal'' i.e., for people of
the same age group, but there are nontrivial interactions between age groups
at some ``distance'' between them (e.g., parents/grand-parents and
children/grand-children); see, e.g.,~\cite{ram1}. There, a more complicated non-monotonic kernel of
interaction across ages may be relevant to include. These are all interesting
possibilities, currently under consideration for future work and will be
reported accordingly in future publications.

\vspace{5mm}

{\it Acknowledgements.} P.G.K. gratefully acknowledges
support through the C3.ai Digital Transformation Institute and also
enlightening
discussions with the PEACoG group (M. Barmann, Q.-Y. Chen,
J. Cuevas-Maraver, Y. Drossinos, G.A. Kevrekidis, Z. Rapti), as well
as with  A. Supernaw.

\bibliographystyle{elsarticle-num}
\bibliography{sir_v2}

\end{document}